# SPECIFIC BEHAVIOR OF ONE CHAOTIC DYNAMICS NEAR THE FINE-STRUCTURE CONSTANT

D.B. Volov, Russia, Samara State University of Transport
18 Pervy Bezymyanny pereulok 443066 Samara RUSSIA.
e-mail: volovdm@mail.ru
www.volovdm1.narod2.ru

**Abstract**

*A chaotic dynamics generalizing the Verhulst, Ricker dynamics and containing a new parameter is introduced. It is established that with the value of this parameter approaching the fine-structure constant $\alpha = 1/137...$ the chaos in the system is considerably weakening. It is shown that there is only one more value of this parameter characterized by ordering initially chaotic dynamics. Two-parameter Mandelbrot/Julia Sets are built for the complex map analogs.*

PACS numbers: 05.45.-a, 05.45.Pq

Key words: one-dimensional discrete Vehulst, Ricker maps, fine-structure constant, bifurcation diagram, fractal.

**Introduction**

Various one-dimensional maps are used to model dynamic processes; the discrete Verhulst-Pearl model is considered the classical one [1]:

$$x_{n+1} \to q x_n (1 - x_n), \tag{1}$$

where $x_n$ is the value at a discrete moment of time $n$, $q$ is the real-valued parameter expressing the rate of x-growth. For some initial conditions and values the model gives negative values $x_{n+1}$. The discrete Ricker model [2] lacks these limitations:

$$x_{n+1} \to q x_n \exp(-x_n), \tag{2}$$

with small $x_n$ leading to (1) when the exponent is expanded as a Taylor series. Model (2) has found wide use in biology and economics. There are its modifications with $\Phi$ exponent when $x_n$ is other than 1.0 [3] − [8].

Let us consider the generalized modification of these models which can be applied to identical particle physics [9] – [14]. The exponent in the Ricker model (2) is written in the denominator and added by small parameter $\alpha$:

$$x_{n+1} \to \frac{q x_n^{\Phi}}{\exp(x_n) + \alpha} \tag{3}$$

($\alpha$, $\Phi$ are real numbers). With $\alpha \to 0$ and $\Phi = 1$ the dynamics (3) becomes (2). Along with characteristic cascades of period doubling bifurcations, periodic windows, etc., the dynamics (3), hereafter called the generalized Verhulst-Ricker-Planck dynamics (VRP), exhibits a number of principally new properties.

**Specific behavior of the unified one-dimensional Verhulst-Ricker-Planck map.**

It turns out that in contrast to the one-dimensional Verhulst-Pearl map (1) and the Ricker discrete model (2), the bifurcation diagram of the generalized Verhulst-Ricker-Planck dynamics (3) demonstrates rather untrivial behavior near certain values of parameter $\alpha$. Thus, with $\Phi = 1$ and the dynamics (3) approaching $\alpha \to -1+0$ it is weakening, whereas with $\alpha = -1$ the chaotic component disappears altogether (fig. 1). As a result, between two values of $x$ there is nothing except oscillations. After the only bifurcation the two branches are developed on $q > 3$ (fig. 1, a, b, c). Thus, the system "is cleaning" to remove chaotic bifurcations and to leave one and only one bifurcation.

With $\Phi_q = 2.633...$ the majorant of the dynamics (3) is exactly equal to number $q$: $\sup(\bar{x}|\Phi_q) \equiv q$. This bifurcation diagram is shown in fig. 2, a. The values of parameter $q$ are set along the x-axis; whereas on the y-axis the values of $x_n$, assuming $x$ on $n \gg 1$, ($\alpha = -1$) are set. Fig. 2, b shows the Lamerey ladder of the attracting 3-cycle with $q = 8.230$; fig 2, c shows the process of oscillations between these three values. Then, with $q$ increasing, the sequence of the shift of the number of values satisfies the Sharkovsky order. The bifurcation diagram displays self similarity (fig. 2) and is a typical example of chaotic behavior. For $\alpha = -1$ the $\Phi_q$ value occurs, if the requirement is met that the right hand side of the recursive formula (3) without $q$ reaches maximum in 1.0, which is equivalent to the following system of transcendental equations:

$$\begin{cases} x/\Phi = 1 - e^{-x} \\ x^\Phi = e^x - 1 \end{cases} ;$$

the solution of the system is provided by $\Phi_q = 2.633...$ with $x = 2.392...$

With $\Phi = -2$ (VRP$^{-2}$-dynamics) and $\alpha \to 1/137... - 0$ the chaotic component of this dynamics

$$x_{n+1} \to \frac{-(q/x_n^2)}{\exp(x_n) + \alpha} \qquad (4)$$

is weakening in the similar way, and in the area of negative $x$ a new characteristic picture of limited bifurcations occurs – «four rats» (fig. 3) [15]. With any initial condition of $x_0$, selected from the basin of attraction, through the finite number of iterations we obtain a strictly ordered picture of the $x$ distribution.

It should be noted that the ordered dynamics (3) exists with $\Phi = -1$, but it is not chaotic. The VRP$^{-2}$ dynamics (4) has an analog in the area of positive $x$:

$$x_{n+1} \to \frac{(q/x_n^2)}{\exp(-x_n) + \alpha} .$$

The VRP$^{-2}$ dynamics takes places on $\alpha \in \sim[-0.430, +0.106]$, $q \in \sim[e^{-4}, e^{+8}]$ and gives the values $|x|$ in the range of $\sim e^{19.5} \approx 3 \cdot 10^8$. With $\alpha = 1/137...$ its whole chaotic part is hidden in the "ears" on the right hand side (fig. 3).

Specific behavior of the VRP$^{-2}$ dynamics is best demonstrated when constructing the logarithmic graph chart (fig. 3).

Thus, the VRP$^{-2}$ dynamics, with the parameter $\alpha$ value approaching the value equal to the value of the dimensionless fine-structure constant, features maximum variety with regard to the number of bifurcations with the minimum degree of chaosticity.

With large positive values of $\alpha > 0.010$ there is no chaos (fig. 4). At first, a "crack" is appearing which undergoes a number of bifurcations in both directions, then the lines are being internally bifurcated (fig. 4, frames 1–3). The prechaotic dynamics is the richest one with the above-mentioned $\alpha = +1/137...$ With even smaller positive $\alpha$ the chaotic dynamics is developing, the zones are intersected resulting in characteristic pictures of bifurcations (fig. 4, frames 5-7). With $\alpha$ approaching 0 on the right, periodic windows are expanding, and with $\alpha = 0$ the dynamics loses its right branch (fig. 4, frame 8). With $\alpha < 0$ and $\alpha$ lowering, the left branch, reserving the chaotic component, is gradually squeezing, exhibiting much greater stability (as compared with the right one) (fig. 4, frames 8-10).

Thus, for the generalized Verhulst-Ricker-Planck dynamics (3) it has been found out that there are two unique values of the dimensionless parameter: $\alpha = -1$ and $\alpha \approx +1/137$. With $\alpha$ approaching the first value, the bifurcation diagram is weakening and degenerating into two branches (fig. 1); approaching the second value it exhibits maximum variety with the minimum degree of chaosticity (fig. 3, 4).

The ratio of distances between the bifurcation points in the logarithmical coordinates (fig. 5) for the VRP$^{-2}$ dynamics both horizontally and vertically is equal to 2: $\frac{x_1}{x_2} = \frac{x_3}{x_4} = \frac{q_1}{q_2} = 2$. The value of $\frac{a}{b}$ of the ratio of gap $a$ at the beginning of bifurcation to gap $b$ at the end of bifurcation is probably equal to $\frac{\pi}{4}$.

The substitution of variables $x \leftarrow x/s$ does not result in new dynamics; it only displays the VRP$^{-2}$ dynamics in relative units:

$$\bar{x}_2(q_2, s_2, \alpha) \equiv \frac{s_2}{s_1} \cdot \bar{x}_1\left(q_2 \cdot \left(\frac{s_1}{s_2}\right)^3, s_1, \alpha\right).$$

The logistic map like (1) in complex numbers can be correlated with the set of the Julia map $z \to z^2+c$. In the same way, the one-dimensional VRP dynamics with complex parameters is correlated with the generalized Julia set. The Mandelbrot set is built according to generally accepted rules.

On the complex plane the inverse VRP$^{-2}$ dynamics

$$z_{n+1} \to \frac{-(c_{ij}/z_n^2)}{\exp(z_n)+\alpha_{ij}} \qquad (5)$$

gives generalized Julia sets which depend both on complex parameter $c_{ij}$, and on complex parameter $\alpha_{ij}$. The generalized Mandelbrot set is built for $z_{ij} \leftarrow 1$.

Fig. 6, *a* shows the Julia set with real numbers $\alpha = 1/137...$ and $c = e = 2.718...$; fig. 6, *b* gives its bottom left quarter. It is possible to rotate this fractal object relative to the visible horizontal axis of symmetry. Provided $\alpha$ and *c* have nonzero complex components, they deform its rotation axis into the hyperbola and break symmetry. With certain parameter values we obtain well known fractals. Thus, with $c \approx 0.10$, $\alpha \approx 0.02$ in the center of the generalized Julia set (5) there appears the well-known Mandelbrot set $z \to z^2 + c$.

As $\alpha$ in (5) is a complex number, in the four-dimensional space $R^3$-$t$ we can display only time-varying "layers" of this mathematical object. The full Mandelbrot set of this object for complex and hypercomplex numbers is no longer expressible in $R^3$-$t$.

**Conclusions**

The main result presented in this paper proves that for the generalized Verhulst-Ricker-Planck dynamics there are two specific values of the dimensionless parameter $\alpha$: –1 and +1/137..., for which a new phenomenon of considerable chaos weakening in the system is observed. Approaching the first value the dynamics is degenerating into two branches, whereas approaching the second value the dynamics exhibits maximum variety with the minimum degree of chaoticity.

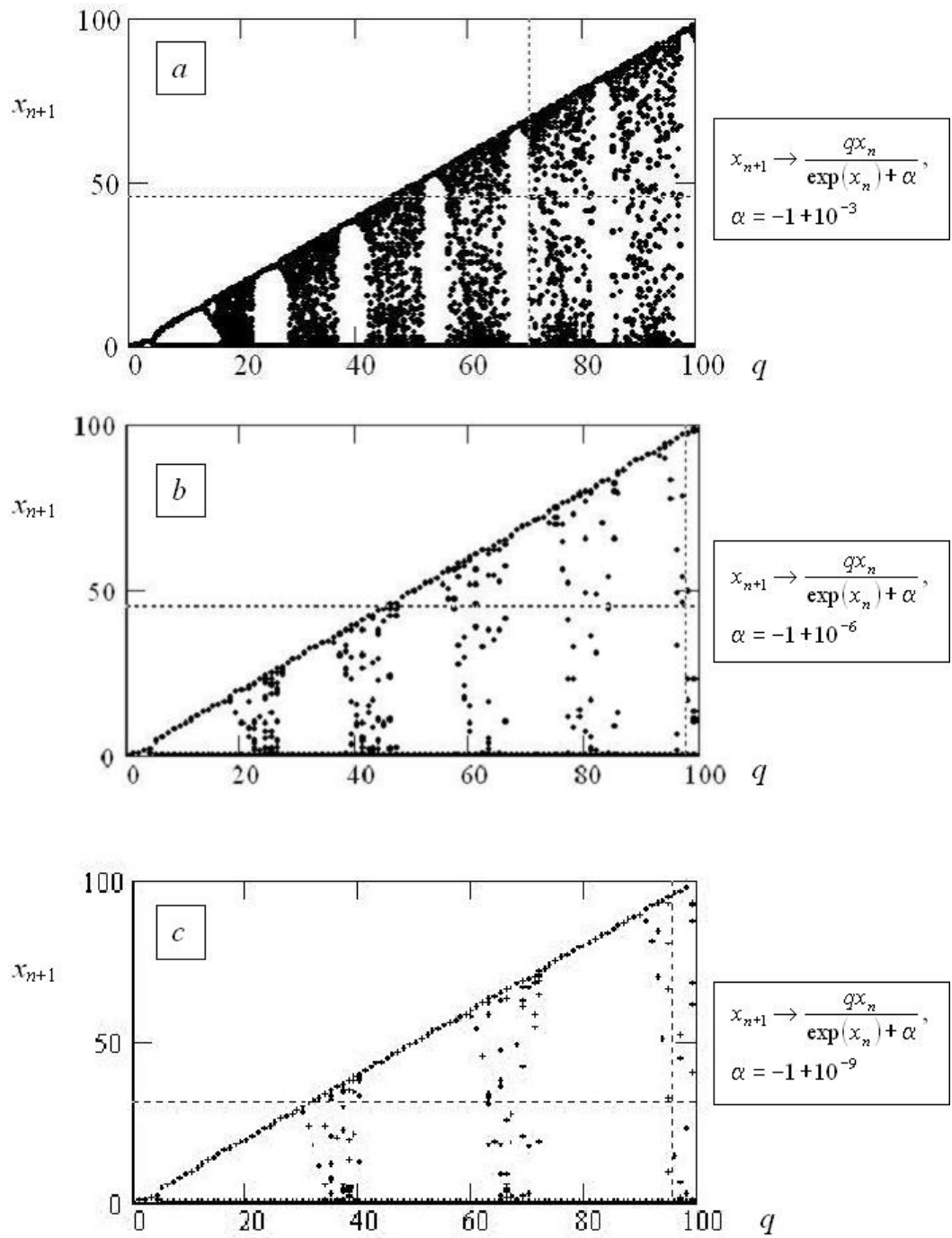

Fig. 1. Dynamics $x_{n+1} \to \dfrac{qx_n}{\exp(x_n)+\alpha}$ "removing" chaos with $\alpha \to -1$.

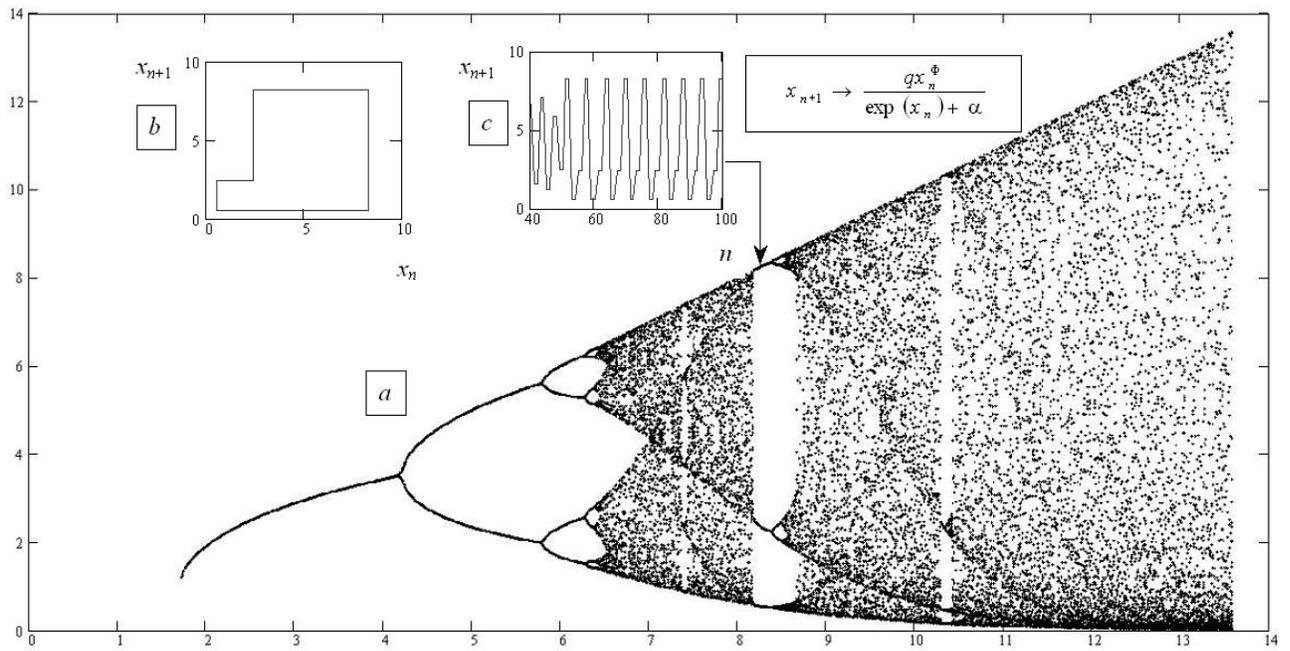

Fig. 2, *a* – the VRP diagram with $\Phi = \Phi_q$, $\alpha = -1$, *b* – an example of the attracting cycle, *c* – its involute.

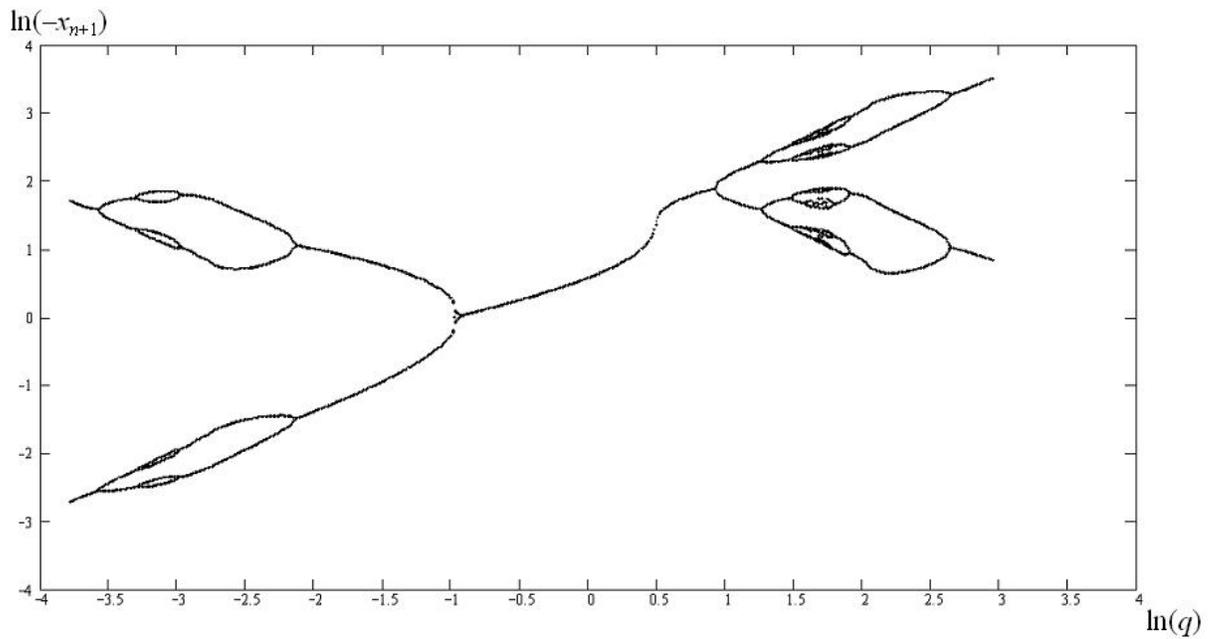

Fig.3. Bifurcation diagram "four rats". Dynamics $x_{n+1} \to \dfrac{-(q/x_n^2)}{\exp(x_n)+\alpha}$, $\alpha = 1/137...$. Disappearing of chaosticity.

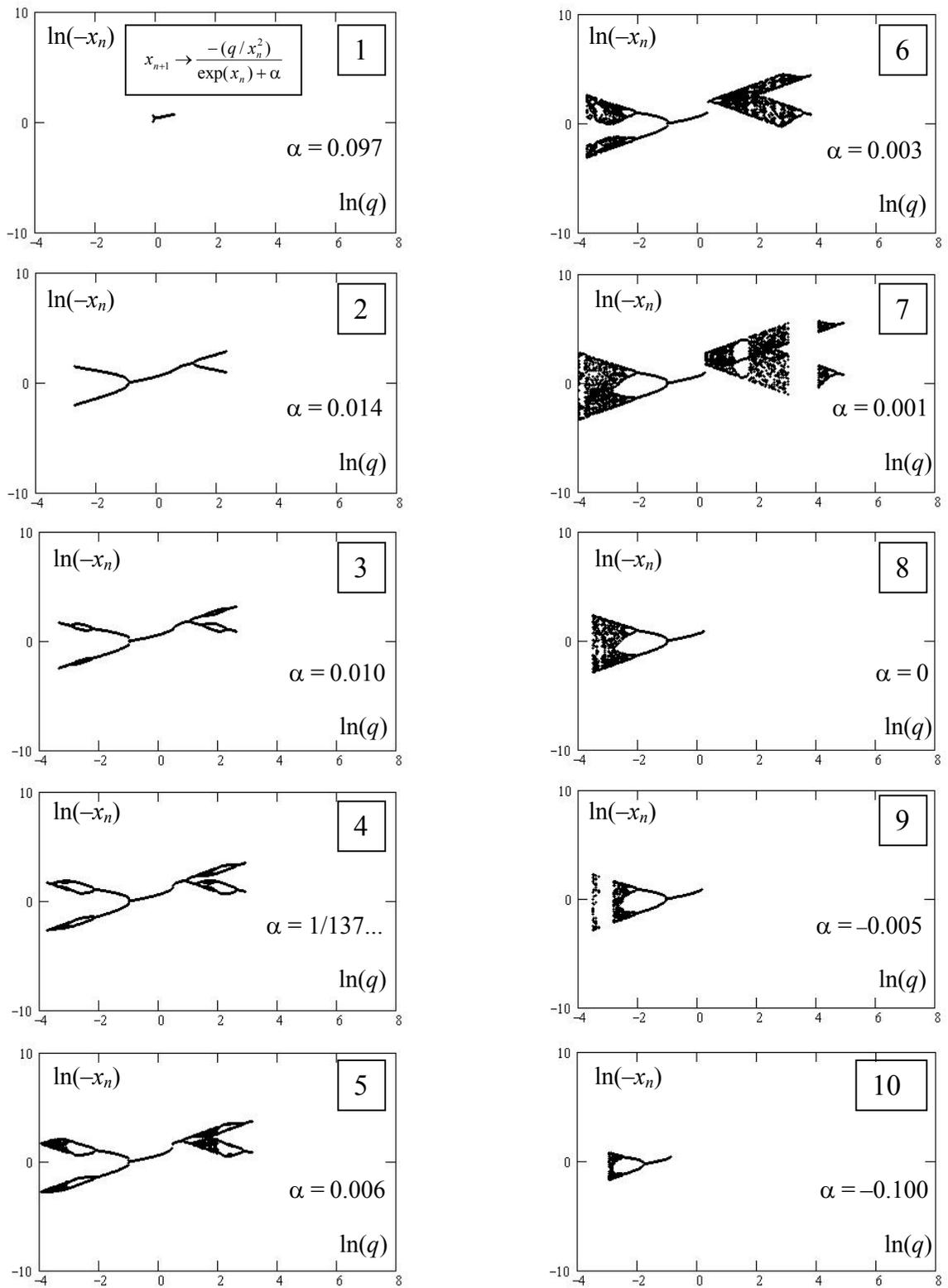

Fig. 4. Dynamics $x_{n+1} \to \dfrac{-(q/x_n^2)}{\exp(x_n)+\alpha}$ with parameter $\alpha$ being changed.

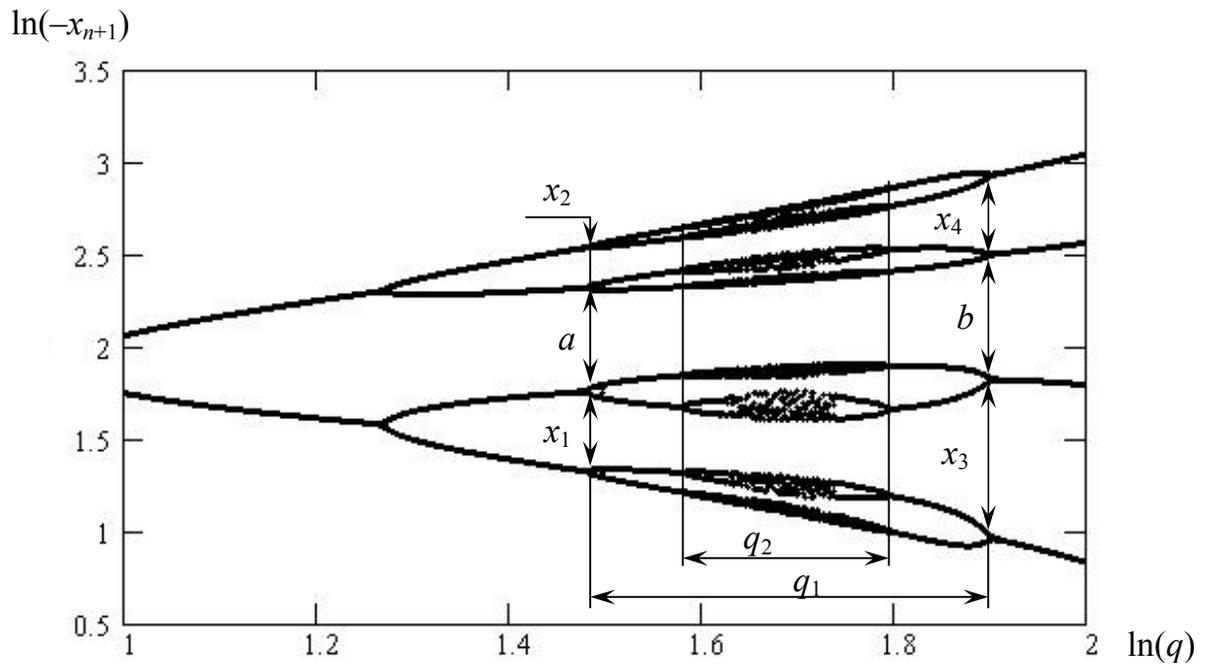

Fig. 5. Balance of bifurcation arms in the VRP$^{-2}$ dynamics with the value of parameter $\alpha$ equal to the fine-structure constant.

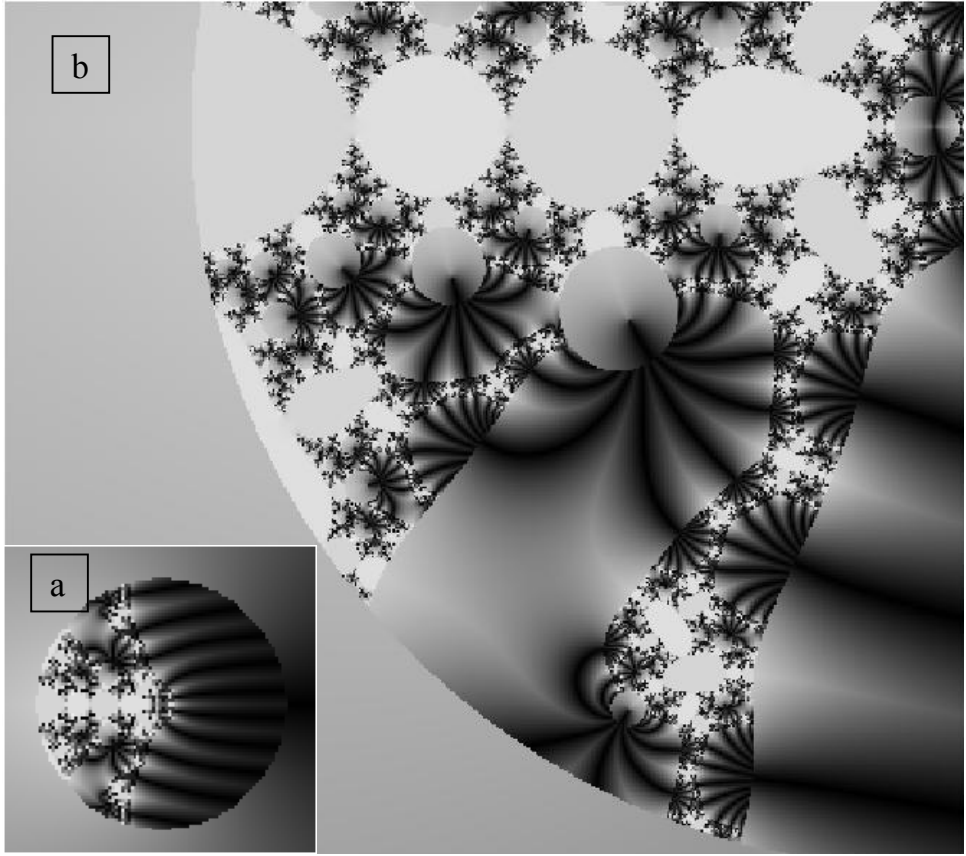

Fig. 6. The generalized Julia set for the reverse dynamics (7), $c = e$, $\alpha = 1/137...$